\documentclass[final,twocolumn,times]{elsarticle}

\usepackage{amssymb}
\usepackage{amsthm}
\usepackage{amsmath}
\usepackage[caption=false,font=normalsize,labelfont=sf,textfont=sf]{subfig}
\usepackage{wrapfig}
\usepackage{graphicx}
\usepackage{color}
\usepackage{multirow}
\usepackage{makecell}
\usepackage[colorlinks]{hyperref}
\usepackage{bbding}
\journal{Arxiv}

\begin{document}
	
	\begin{frontmatter}
		\title{EDMAE: An Efficient Decoupled Masked Autoencoder for Standard View Identification in Pediatric Echocardiography\tnoteref{support}}
		\tnotetext[support]{This project was partially supported by the Natural Science Foundation of China (Grant No.61975056, 661801288), the Shanghai Natural Science Foundation (Grant No.19ZR1416000), the Shanghai Municipal Health Commission (Project No.20194Y0321) and the Minhang District Public Health Excellent Youth Talent Training Project (Project No.2020FM29).}
		
		\author[scmc,sercip,ecnu]{Yiman Liu\fnref{ce}}
		\ead{LiuyimanSCMC@163.com}
		\affiliation[scmc]{organization={Department of Pediatric Cardiology, Shanghai Children's Medical Center, School of Medicine, Shanghai Jiao Tong University},
			city={Shanghai},
			postcode={200127},
			country={China}}
		\affiliation[sercip]{organization={Shanghai Engineering Research Center of Intelligence Pediatrics (SERCIP)},
			city={Shanghai},
			postcode={200127},
			country={China}}
		\affiliation[ecnu]{organization={Shanghai Key Laboratory of Multidimensional Information Processing, School of Communication and Electronic Engineering, East China Normal University},
			city={Shanghai},
			postcode={200241},
			country={China}}
		
		\author[usst]{Xiaoxiang Han\fnref{ce}}
		\ead{gtlinyer@163.com}
		\fntext[ce]{These authors contributed equally to the work.}
		\affiliation[usst]{organization={School of Health Science and Engineering, University of Shanghai for Science and Technology},
			city={Shanghai},
			postcode={200093},
			country={China}}
		
		\author[mdcdcp]{Tongtong Liang\fnref{ce}}
		\ead{zhimaliang@163.com}
		\affiliation[mdcdcp]{organization={Shanghai Minhang Center for Disease Control and Prevention},
			city={Shanghai},
			postcode={201101},
			country={China}}
		
		\author[sercip]{Bin Dong}
		\ead{dongbin@scmc.com.cn}
		
		\author[sercip]{Jiajun Yuan}
		\ead{y680317@163.com}
		
		\author[ecnu]{Menghan Hu}
		\ead{mhhu@ce.ecnu.edu.cn}
		
		\author[sumhs]{Qiaohong Liu\corref{cor}}
		\ead{hqllqh@163.com}
		\cortext[cor]{Authors are equally corresponded.}
		\affiliation[sumhs]{organization={School of Medical Instruments, Shanghai University of Medicine and Health Sciences},
			city={Shanghai},
			postcode={201318},
			country={China}}
		
		\author[ecnu]{Jiangang Chen\corref{cor}}
		\ead{jgchen@cee.ecnu.edu.cn}
		
		\author[ecnu]{Qingli Li\corref{cor}}
		\ead{qlli@cs.ecnu.edu.cn}
		
		\author[scmc,sercip]{Yuqi Zhang\corref{cor}}
		\ead{changyuqi@hotmail.com}
		
		\begin{abstract}
			This paper introduces the Efficient Decoupled Masked Autoencoder (EDMAE), a novel self-supervised method for recognizing standard views in pediatric echocardiography. EDMAE introduces a new proxy task based on the encoder-decoder structure. The EDMAE encoder is composed of a teacher and a student encoder. The teacher encoder extracts the potential representation of the masked image blocks, while the student encoder extracts the potential representation of the visible image blocks. The loss is calculated between the feature maps output by the two encoders to ensure consistency in the latent representations they extract. EDMAE uses pure convolution operations instead of the ViT structure in the MAE encoder. This improves training efficiency and convergence speed. EDMAE is pre-trained on a large-scale private dataset of pediatric echocardiography using self-supervised learning, and then fine-tuned for standard view recognition. The proposed method achieves high classification accuracy in 27 standard views of pediatric echocardiography. To further verify the effectiveness of the proposed method, the authors perform another downstream task of cardiac ultrasound segmentation on the public dataset CAMUS. The experimental results demonstrate that the proposed method outperforms some popular supervised and recent self-supervised methods, and is more competitive on different downstream tasks.
			
		\end{abstract}
		
		
	\end{frontmatter}
	
	\section{Introduction}
	Congenital heart diseases (CHDs) are the most prevalent types of birth defects, affecting approximately 0.9\% of live births. Unfortunately, they also represent the primary cause of death among children between the ages of 0 and 5~\cite{zhao2019prevalence}. Each year, around 100,000-150,000 newborns in China are diagnosed with CHD, and the incidence of CHD has been steadily rising since the full implementation of the second-child policy in 2016. Therefore, early and precise diagnosis of CHD holds significant clinical importance.
	
	Transthoracic echocardiography (TTE) is a cost-effective, non-invasive, and radiation-free imaging technique that enables real-time and dynamic visualization of the heart. TTE has become an essential tool for the diagnosis and treatment of CHD due to its ability to rapidly detect various cardiac abnormalities~\cite{douglas2011accf}. TTE involves standard view acquisition, dynamic image scanning, and measurement. Among these steps, the precise acquisition of standard views is a prerequisite for subsequent measurement of biological features and the final diagnosis of CHD.
	
	Nevertheless, the complex and variable anatomy and spatial configuration of congenital heart disease make accurate diagnosis through TTE challenging and time-consuming, necessitating experienced cardiac specialists to carefully interpret each ultrasound image. The American Society of Echocardiography recommends the use of standard imaging techniques for 2D, M-mode, and color Doppler echocardiography~\cite{lopez2010recommendations}. This entails acquiring images following a reproducible protocol. Specifically, the acquisition of images in a particular view is necessary to facilitate the measurement of specific structures and minimize inter- and intra-observer variability~\cite{burgos2020evaluation}.
	
	Hence, the application of deep learning techniques for the automatic intelligent recognition of standard views in pediatric echocardiography becomes imperative. This approach not only forms the basis for intelligent CHD diagnosis but also offers standardized training for primary cardiac ultrasonographers to perform view sweeping, thereby providing valuable clinical applications.
	
	Deep learning has rapidly been applied to the medical field due to the continuous development of artificial intelligence. UNet~\cite{ronneberger2015u}, for instance, has been proposed for medical image segmentation. Deep learning being data-driven requires a large amount of annotated data to fit the target function. Annotating a large amount of data is expensive, particularly in the medical field where the number of images is small, and accurate data annotation is challenging. This study collected a large number of children's echocardiograms, and annotating each image was costly and time-consuming. Furthermore, although pre-training on a large-scale dataset can improve the network's performance to some degree, natural image to medical image transfer often yields poor results. Self-supervised learning has become increasingly popular in recent years because it can reduce the cost of annotating large-scale datasets by using custom pseudo-labeling to supervise training and learned latent representation for multiple downstream tasks~\cite{jaiswal2020survey}. The masked autoencoder, as a powerful self-supervised method, has recently been rapidly applied to medical image analysis~\cite{zhou2022self,tian2022unsupervised,xiao2023delving,xu2022swin,chen2022multi}. Autoencoders were introduced into medical image analysis by Zhou et al.~\cite{zhou2022self}, and they were verified on multiple medical datasets and tasks. Tian et al..~\cite{tian2022unsupervised} used a memory-enhanced multi-level cross-attention masked autoencoder for unsupervised anomaly detection in medical images. Xiao et al.~\cite{xiao2023delving} conducted in-depth research on the masked autoencoder for multi-label thoracic disease classification and achieved advanced performance on chest X-ray images. Additionally, some researchers~\cite{xu2022swin} replaced the ViT~\cite{dosovitskiy2020image} used by MAE~\cite{he2022masked} with Swin Transformer~\cite{liu2021swin} to adapt to small medical datasets, while others~\cite{chen2022multi} applied the masked autoencoder to medical multimodal data.
	
	Self-supervised pre-training for images involves learning from degradation, which entails removing specific information from the image signal and requiring the algorithm to restore it. However, this degradation-based method faces a significant bottleneck, which is the conflict between degradation intensity and semantic consistency. Visual representation learning relies wholly on degradation since there is no supervised signal, and the degradation must be strong enough. Nonetheless, when the degradation is strong enough, it is not guaranteed that the images before and after degradation have semantic consistency. To address this issue, we propose an efficient decoupled masked autoencoder (EDMAE). The EDMAE has two identical encoders: the teacher encoder, which takes visible image blocks as input and can backpropagate to update weights, and the student encoder, which takes a mask as input and cannot backpropagate, updating weights from the teacher encoder. Their latent representations are the feature maps they output, and their alignment is maintained by calculating the loss between the representations of visible image blocks and mask image blocks. This approach ensures the encoder's representations from any part of the image are consistent, which can compel the encoder to learn more latent representation information. Consequently, the encoder is decoupled from the decoder, preventing the decoder from learning representation information and allowing it to concentrate on the reconstruction task. Additionally, our proposed method relies on pure convolutional operations~\cite{huang2017densely}, which are lighter and have faster training and convergence speeds than the ViT used by the MAE and BEiT~\cite{bao2021beit}. The proposed method, similar to MAE, utilizes asymmetric encoders and decoders to decrease the time and memory usage required for pre-training.
	
	We pre-trained the proposed method on a large-scale unlabeled dataset of pediatric cardiac ultrasound images constructed in this study. We then validated it on private pediatric cardiac ultrasound standard view recognition. Furthermore, we conducted experiments on the public dataset CAMUS to verify that the proposed method can extract effective representations from the pre-trained ultrasound cardiac dataset.
	
	The paper is structured as follows: The Introduction section provides the research background, motivation, research questions, and objectives of the study. The Related Works section reviews the existing literature on self-supervised learning and its applications in medical image analysis. The Proposed Method section describes our proposed method, including its main architecture, the self-supervised pre-training process, and the downstream task fine-tuning process. The Experiment section presents the results of our comparative and ablation experiments. The Discussion section evaluates the advantages and disadvantages of our model and outlines future plans. Finally, the Conclusion section summarizes our proposed method and its performance results.
	
	The main contributions of this paper are as follows.
	\begin{enumerate}
		\item We propose an efficient decoupled mask autoencoder (EDMAE) that decouples the encoder and decoder. This enforces the encoder to learn high-quality latent representations. 
		
		\item The proposed method uses an asymmetric encoder-decoder structure. DenseNet is used as the encoder, while a lightweight CNN is used as the decoder. This approach enhances the method's efficiency, with lower computational costs and faster convergence speed. 
		
		\item We utilized the proposed method for self-supervised pre-training on a large-scale private dataset of children's hearts that we collected. We then fine-tuned it on two downstream tasks. The experimental results demonstrate the superiority of the proposed method.
	\end{enumerate}
	
	\section{Related Works}
	\subsection{Self-supervised learning}
	Self-supervised learning can be categorized into two main types: generative and contrastive. Contrastive learning (CL) is a discriminative method that brings similar samples closer together while pushing different samples farther apart. In 2020, the introduction of MoCo~\cite{he2020momentum} brought contrastive learning to a new stage by using a dynamic dictionary library, avoiding the memory bottleneck problem faced by SimCLR~\cite{chen2020simple}. MoCo achieved accuracy levels close to those obtained through supervised training.
	
	Generative learning is another form of self-supervised learning. Since the introduction of Generative Adversarial Networks (GANs)~\cite{goodfellow2020generative} in 2014, generative models have made significant progress. Recently, Masked Image Modeling (MIM) has become a popular generative self-supervised algorithm with the introduction of MAE, SimMIM~\cite{xie2111simple}, and BEiT. These methods learn feature representations by compressing input data into an encoding and then reconstructing the input. Recently, several works have been proposed to improve this method, such as CAE~\cite{chen2022context} and TACO~\cite{fu2022contextual}.
	
	\subsection{Self-supervised learning in medical image analysis}
	In the field of medical image analysis, data with high-quality annotations are very scarce. Therefore, self-supervised methods have been quickly introduced in this area. Sowrirajan et al.~\cite{sowrirajan2021moco} used the contrastive self-supervised method MoCo for self-supervised pre-training on a chest X-ray dataset, and then fine-tuned on CheXpert with labeled data. They found that self-supervised pre-training on medical datasets was better than supervised ImageNet pre-trained models. Navarro et al.~\cite{navarro2022self}'s work showed that self-supervised methods outperform previous supervised algorithms in multi-organ segmentation tasks.
	
	Additionally, generative self-supervised algorithms have been proposed for medical image analysis. Ly et al. \cite{ly2022student} proposed the Double Loss Adaptive Masked Autoencoder (DAMA) for multi-immunofluorescence brain image analysis, and their method achieved excellent results on multiple tasks. Quan et al.~\cite{quan2022global} proposed a Global Contrastive Masked Autoencoder for processing pathological images, which achieved competitive results compared to other methods. Furthermore, there are many new research findings~\cite{zhou2022self,tian2022unsupervised,xiao2023delving,xu2022swin,chen2022multi}.
	
	\subsection{Autoencoder}
	Autoencoder (AE) is an self-supervised learning algorithm that learns representations of input information by using the input itself as the learning target~\cite{bengio2013representation}. Classic autoencoders include PCA and k-means~\cite{hinton1993autoencoders}. Since the introduction of Masked Autoencoder (MAE) and BEiT in 2021, autoencoders have become increasingly popular in computer vision self-supervised learning. Recently, they have been increasingly applied in medical image analysis~\cite{xiao2023delving,ly2022student,quan2022global,jiang2022self}.
	
	\section{Proposed Method}
	\subsection{Masked Autoencoder}
	The MAE algorithm employs a random masking technique to obscure certain patches of the input image, which it then reconstructs by filling in the missing pixels. This approach is based on two fundamental design principles: (1) an asymmetric encoder-decoder architecture that handles visible patches differently than mask tokens. The encoder encodes only visible patches and disregards mask tokens, whereas the decoder utilizes the encoder's output (i.e., a latent representation) and mask tokens to reconstruct the image. (2) Using a higher mask ratio has demonstrated promising outcomes. Specifically, a mask rate of 75\% has been shown to produce favorable results. The MAE algorithm operates in several steps. First, it divides the input image into patches and applies a masking operation. Next, it feeds only the vSisible patches into the encoder, along with the mask tokens. The encoder's output and the mask tokens are then used as input to the lightweight decoder, which reconstructs the entire image. The loss function used is the mean squared error (MSE) loss, which is only computed for the masked patches. MAE has demonstrated robust transferability and achieved the highest accuracy of 87.8\% on the ImageNet-1K dataset. Moreover, due to its simplicity, it is highly scalable, making it an attractive option for large-scale image processing applications.
	
	In an asymmetric encoder-decoder architecture, the encoder and decoder have different numbers of layers or different numbers of neurons in each layer. The encoder of the proposed method is DenseNet, while the decoder adopts a lightweight CNN. This can provide several benefits over a symmetric architecture where the encoder and decoder have the same structure: (1) It can help to reduce the computational complexity of the network. By using a smaller decoder than encoder, the network can be trained to extract the most important features of the input data while discarding less important information. This can lead to faster training times and better performance on test data. (2) It can help to reduce the computational complexity of the network. By using a smaller decoder than encoder, the network can be trained to extract the most important features of the input data while discarding less important information. This can lead to faster training times and better performance on test data. (3) By learning a more complex representation of the input data, the network is better able to generalize to new and unseen data. This can lead to better performance on many tasks. (4) By having a larger encoder network and a smaller decoder network, the network is able to learn a more complex representation of the input data. This can lead to better performance on tasks such as image or speech recognition. In addition, the anatomical structures of the children's echocardiograms we collected are relatively fixed, which means that these images have high redundancy. Therefore, using a higher mask rate for the images can enable the model to learn better potential representations.
	
	Since the MAE model did not completely separate the encoder and decoder, the decoder in MAE still learned latent representations. Therefore, the proposed DEMAE model attempts to decouple the encoder and decoder by using two identical encoders. One of the encoders, called the teacher encoder, takes visible images as input and can be back-propagated to update weights. The other encoder, called the student encoder, takes masks as input and cannot be back-propagated, with weights updated from the teacher encoder. The feature maps output by the encoders are their latent representations, and their consistency is maintained by calculating the loss between the representations of visible images and masks. Therefore, the encoder can learn more latent representations from any part of the image. In addition, unlike MAE, the proposed DEMAE model is based on pure convolutional operations, which have faster training and convergence speeds. The convolutional neural network used is DenseNet, which has strong fitting ability as well as appropriate parameter and computational complexity.

	\begin{figure}[!t]
		\includegraphics[width=1\linewidth]{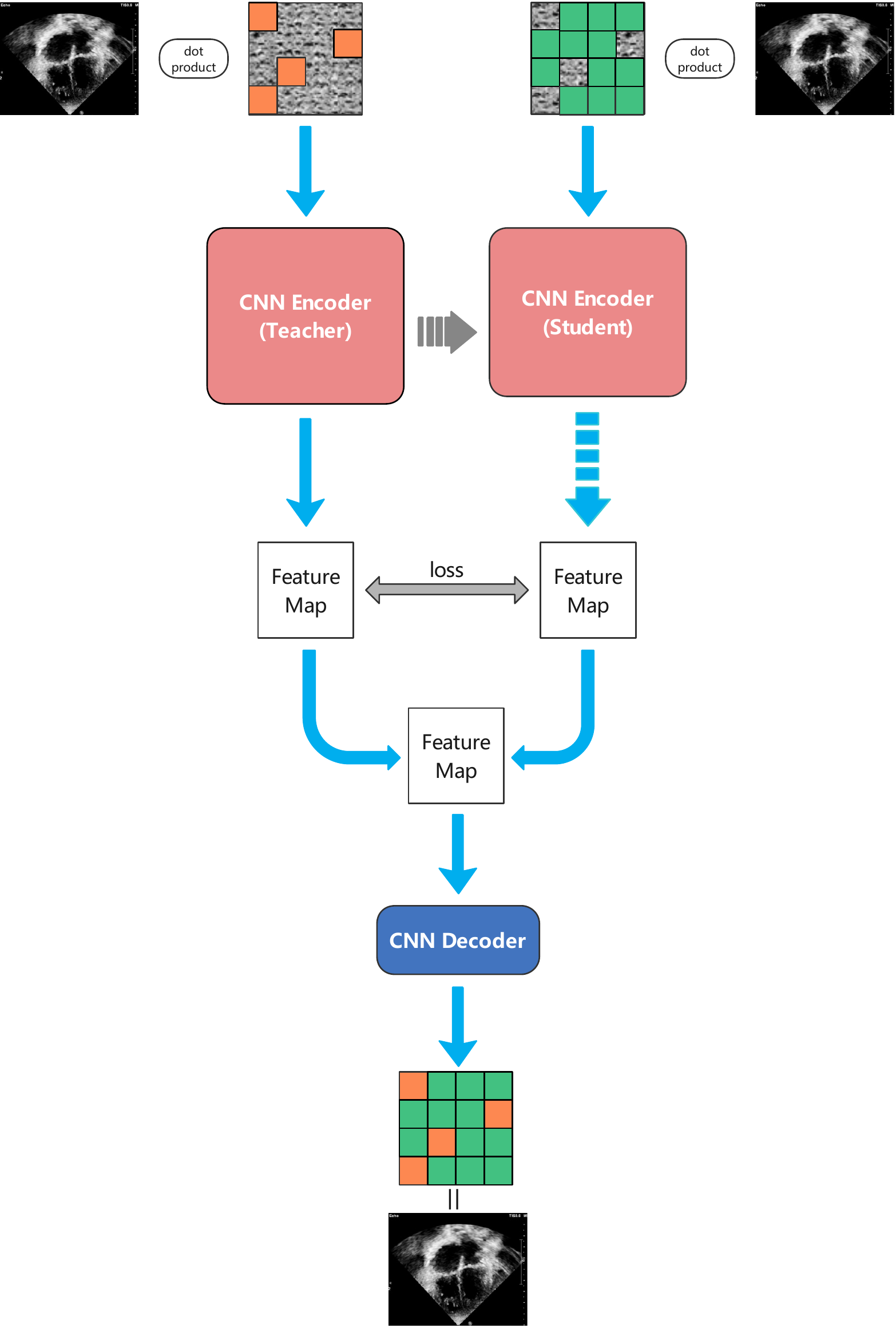}
		\caption{The overall architecture of EDMAE.}
		\label{EDMAE}
	\end{figure}
	
	\subsection{Overall structure of EDMAE}
	The proposed method consists of two inputs, namely the visible and invisible parts of the input image, as shown in Fig.\ref{EDMAE}. A convolutional neural network in a proxy task predicts the invisible part from the visible part, forcing the encoder to learn the latent representation of the image. The proposed method uses two encoders with convolutional neural networks called DenseNet~\cite{huang2017densely}. One encoder is updated through backpropagation and is called the teacher encoder, while the other encoder's backpropagation is blocked and cannot update its weights. It is called the student encoder, which shares weights with the teacher encoder. The decoder and encoder use the same network to predict masked image blocks. The proposed method computes losses in two places, one is between the feature maps output by the two encoders, and the other is between the reconstructed image output by the decoder and the original image. In the proposed method, the momentum update rule is used to update the weights of the student encoder, which is given by the following formula:
	\begin{equation}
		Ps = Ps * m + Pt * (1 - m)
	\end{equation}
	where Ps represents the weight of the student encoder, Pt represents the weight of the teacher encoder, and m represents the momentum.
	
	\subsection{Self-supervised pretraining}
	
	\begin{figure}[!t]
		\centering
		\subfloat[original]{
			\includegraphics[width=0.5\linewidth]{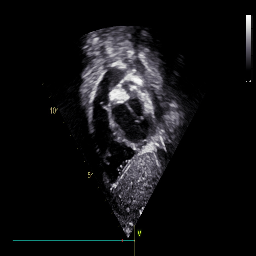}
			\label{original}
		}
		\subfloat[masked]{
			\includegraphics[width=0.5\linewidth]{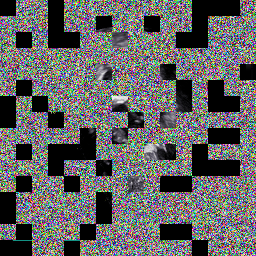}
			\label{masked}
		}\\
		\subfloat[unmasked]{
			\includegraphics[width=0.5\linewidth]{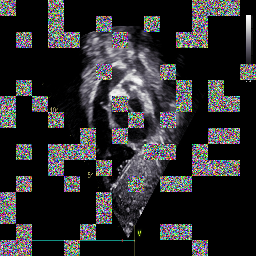}
			\label{unmasked}
		}
		\subfloat[reconstruction]{
			\includegraphics[width=0.5\linewidth]{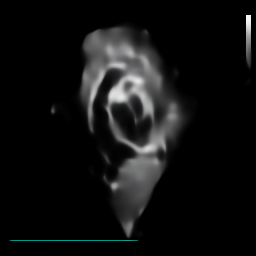}
			\label{reconstruction}
		}
		\caption{The original image, the image masked by 75\%, the unmasked part of the image, and the reconstructed image.}
		\label{image_examples}
	\end{figure}
	
	The proposed self-supervised pretraining method is formally an optimization problem, which is a task of solving static linear inverse problems through a deep neural network. $z\in \mathbb{R}^L$ serves as input, and by optimizing the parameters $\theta$ of the untrained neural network $f_\theta$, it generates an output $f_\theta(z)$ that is consistent with the measurement values $y\in \mathbb{R}^M$. 
	\begin{equation}
		\theta^\ast = \mathop{argmin}\limits_{\theta}{{\left \| y - F(f_\theta(z)) \right \|}_2^2}
	\end{equation}
	where $F\in \mathbb{R}^{M\times N}$ is the forward model. In this paper, y is the masked image, and F is the masking operation. The output of the optimized network $x^*=f_{\theta^*}(z)$ produces remarkably high-quality reconstructed images. 
	
	The proposed self-supervised training method follows the workflow described below. Initially, visible image blocks are fed into the encoder to extract their representations. Next, predictions are made in the encoding representation space, ensuring that the masked image blocks' representations are consistent with those predicted from the visible image blocks. Finally, the decoder takes in the representations of the masked image blocks to predict the masked image blocks. Since previous research~\cite{he2022masked} has shown that a 75\% masking rate produces optimal representations in autoencoders, this paper will use a 75\% masking rate by default. As shown in Fig.\ref{image_examples}., the original image, the image masked by 75\%, the unmasked part of the image, and the reconstructed image are shown in the figure.
	
	The loss calculated between the feature maps output by the two encoders is called feature alignment. The advantage of this approach is that it can ensure that the representation of the mask image block is consistent with the representation obtained from the prediction of the visible image block, ensuring that the image before and after degradation has semantic consistency. It can be represented by the following expression:
	
	\begin{equation}
		y_1=F(x_m)
	\end{equation}
	
	\begin{equation}
		y_2=F^*(x_{um})
	\end{equation}
	
	\begin{equation}
		loss=MSE(y_1,y_2)=\frac{1}{M}\sum_{0}^{m}{(y_1-y_2)^2}
	\end{equation}
	
	Among them, $x_m$ represents the masked image, $x_{um}$ represents the unmasked image, and MSE represents the mean squared error loss function (MSE Loss, L2 Loss).
	
	\subsection{Downstream task}
	
	\begin{table*}[ht]
		\begin{center}
			\caption{Comparative Experiments on Private Datasets.}
			\label{comparison_on_private_dataset}
			\setlength{\tabcolsep}{0.5mm}{
				\begin{tabular}{ l | c | c  c  c | c  c  c }
					\hline
					Method & Overall Accuracy (\%) & Mean Precision (\%) & Mean Recall (\%) & Mean Specificity (\%) & Mean F1 (\%)\\
					\hline
					mobileNetV3-L~\cite{howard2019searching} & 98.17 & 89.68 & 90.10 & 99.57 & 89.06\\
					ResNet50~\cite{he2016deep} & 98.34 & 91.80 & 92.30 & 99.65 & 91.58\\
					Swin-T-B~\cite{xu2022swin} & 98.35 & 92.12 & 92.46 & 99.65 & 91.68\\
					DenseNet121~\cite{huang2017densely} & 98.35 & 92.29 & 92.66 & 99.66 & 91.73\\
					MoCoV2~\cite{chen2020improved} & 98.36 & 92.34 & 92.89 & 99.66 & 91.85\\
					MAE~\cite{he2022masked} & 98.38 & 92.77 & 93.03 & 99.68 & 92.54\\
					ConvMAE~\cite{gao2022convmae} & 98.45 & 92.82 & 93.51 & 99.70 & 92.97\\
					ours & \textbf{98.48} & \textbf{93.20} & \textbf{94.62} & \textbf{99.73} & \textbf{93.63}\\
					\hline
			\end{tabular}}
		\end{center}
	\end{table*}
	
	After completing self-supervised pre-training, all that is needed is to replace the decoder of the proposed method with a task-specific head that caters to the downstream task's characteristics.
	
	For the task of standard view recognition in pediatric cardiac ultrasound, the labels consist of multiple fixed categories, making it an image classification task. Therefore, the decoder needs to be replaced with a linear layer, and cross-entropy is used as the loss function to fine-tune the entire network.
	
	For the task of cardiac ultrasound segmentation, a segmentation task head is required. In this paper, we use the decoder of our own implementation of DenseUNet as the segmentation head. Specifically, the feature maps output by the encoder are used as the input to the segmentation task head, which outputs the segmentation results. The Focal loss is used to compute the loss between the segmentation results and the ground truth labels. 
	
	\section{Experiment}
	
	\subsection{Dataset}
	Our dataset is divided into a private dataset of children's cardiac ultrasound views and a public dataset CAMUS~\cite{leclerc2019deep}. We collected a private dataset from the Department of Pediatric Cardiology, Shanghai Children's Medical Center, School of Medicine, Shanghai Jiao tong University, Shanghai, China. Our study has been approved by the ethics committee of the center (Approval No.: SCMCIRB-K2022183-1). The private children's cardiac ultrasound view data is divided into two parts, one of which has 17,755 unlabeled children's cardiac ultrasound view data for self-supervised pre-training. The other part is the labeled children's echocardiography standard view data with 1026 images for fine-tuning. The data used for fine-tuning includes 616 training sets, 205 validation sets and 205 test sets, which cover 27 standard views of children's echocardiography, 1 other blood flow spectrum and 1 other views. The CAMUS dataset contains two-chamber and four-chamber acquisitions from 500 patients, as well as reference measurements from one cardiologist for the full dataset and three cardiologists for 50 patients. 
	
	\subsection{Training Details}
	We designed our model based on the machine learning framework PyTorch1.12.1 using Python3.8. In particular, we also use PyTorch-Lightning1.6.5, an efficient and convenient framework based on PyTorch. In addition, some of our comparison experiments and ablation experiments use the backbone network provided in Torchvision0.13.1.
	
	We trained the proposed model on a GPU server with an Intel Core i9-10900X CPU, two 10GB Nvidia RTX3080 GPUs, 32GB RAM, and 20GB VRAM. 
	
	We set the batch size of data according to different networks to ensure maximum memory utilization. The number of threads of the data reading program is 16. The initial learning rate is 1e-3. The learning rate dynamic adjustment strategy is ReduceLROnPlateau. The optimizer is AdamW~\cite{loshchilov2017fixing}. The training epoch number is 100. Train with automatic mixed precision.
	
	The loss function used for pre-training is the mean square error (MSE) loss function. The loss function for downstream classification tasks is the cross-entropy loss function. The loss function for downstream segmentation tasks is Focal Loss~\cite{lin2017focal}, which can reduce the weight of easily classified samples and increase the weight of difficult-to-classify samples. Its formula is as follows:
	
	\begin{equation}
		\label{FocalLoss}
		FL(p_t) = -\alpha{_t}(1-p_t)^\gamma\log{(p_t)}
	\end{equation}
	
	p$\in$[0,1] is the model's estimated probability of the labeled class, $\gamma$ is an adjustable focusing parameter, and $\alpha$ is a balancing parameter. We set $\gamma$ to 2 and $\alpha$ to 0.25.
	
	\subsection{Evaluation Metrics}
	To evaluate the performance of the proposed EDMAE, we use some commonly used metrics to assess the accuracy of the model. For classification tasks, we use Overall Accuracy (OA), Precision, Recall, Specificity, and F1-Score (F1) . These evaluation metrics are calculated based on a confusion matrix, where TP represents the number of True Positive samples, TN represents the number of True Negative samples, FP represents the number of False Positive samples, and FN represents the number of False Negative samples. 
	
	Overall Accuracy (OA) is used to measure the overall accuracy of the model's predicted results:
	
	\begin{equation}
		\label{OA}
		OA = \frac{TP+TN}{TP+TN+FP+FN}
	\end{equation}
	
	F1 Score represents a comprehensive consideration of Precision and Recall:
	
	\begin{equation}
		\label{Precision}
		Precision = \frac{TP}{TP+FP}
	\end{equation}
	\begin{equation}
		\label{Specificity}
		Specificity = \frac{TN}{FP+TN}
	\end{equation}
	\begin{equation}
		\label{Recall}
		Recall = \frac{TP}{TP+FN}
	\end{equation}
	\begin{equation}
		\label{F1}
		F1 = 2\frac{Precision\times{Recall}}{Precision+Recall} = \frac{2TP}{2TP+FP+FN}
	\end{equation}
	
	For the task of cardiac ultrasound segmentation, we adopt three metrics: Dice coefficient (DC), Hausdorff distance (HD), and area under the curve (AUC).
	
	\begin{equation}
		\label{DC}
		DC=\frac{2\times \left| A\cap B \right|}{\left| A \right| + \left| B \right|}=\frac{2TP}{2TP+FN+FP}
	\end{equation}
	
	\begin{equation}
		\label{HD}
		\begin{aligned}
			& HD=\max{\{d_{AB},d_{BA}\}} \\
			& =\max{\{\mathop{\max}\limits_{a\in{A}}\mathop{\min}\limits_{b\in{B}}{d_{(a,b)}}, \mathop{\max}\limits_{b\in{B}}\mathop{\min}\limits_{a\in{A}}{d_{(a,b)}}\}}
		\end{aligned}
	\end{equation}
	
	\subsection{Experimental results on the private dataset}
	
	\begin{table*}[!t]
		\begin{center}
			\caption{Experimental results on the public dataset CAMUS.}
			\label{results_on_the_public_dataset_CAMUS}
			\setlength{\tabcolsep}{8mm}{
				\begin{tabular}{ l | c  c  c }
					\hline
					Method & DC (\%) & HD (mm) & AUC (\%)\\
					\hline
					Joint-net~\cite{ta2020semi} & 91.05 $\pm$ 0.27 & 3.41 $\pm$ 0.86 & 97.14 $\pm$ 0.25\\
					DenseUNet~\cite{huang2017densely} & 91.88 $\pm$ 0.26 & 3.34 $\pm$ 0.82 & 97.26 $\pm$ 0.24\\
					TransUNet~\cite{chen2021transunet} & 91.89 $\pm$ 0.38 & 3.25 $\pm$ 1.01 & 97.39 $\pm$ 0.24\\
					MFP-Net~\cite{moradi2019mfp} & 92.23 $\pm$ 0.29 & 3.40 $\pm$ 0.97 & 97.28 $\pm$ 0.23\\
					PLANet~\cite{liu2021deep} & 92.61 $\pm$ 0.40 & 3.10 $\pm$ 0.93 & 97.58 $\pm$ 0.23\\
					ours & \textbf{93.09 $\pm$ 0.22} & \textbf{3.02 $\pm$ 0.81} & \textbf{97.84 $\pm$ 0.22}\\
					\hline 
			\end{tabular}}
		\end{center}
	\end{table*}
	
	The proposed method was evaluated on a private dataset created for this study. We compared the proposed method with several mainstream classification networks, namely MobileNetV3-large~\cite{howard2019searching}, ResNet50~\cite{he2016deep}, Swin-Transformer-base~\cite{xu2022swin}, and DenseNet121~\cite{huang2017densely}. These networks were selected from the TorchVision built-in model module and were pre-trained on ImageNet-1k. Additionally, we compared our method with some recent self-supervised methods, including MoCoV2~\cite{chen2020improved}, MAE~\cite{he2022masked}, and ConvMAE~\cite{gao2022convmae}. Table~\ref{comparison_on_private_dataset} shows that the proposed method outperforms other methods in the majority of metrics. The F1 Score is 0.66\% higher than ConvMAE, Precision is 0.38\% higher than ConvMAE, Recall is 1.11\% higher than ConvMAE, and Specificity is 0.03\% higher than ConvMAE. Overall, the proposed method is highly competitive.
	
\begin{figure}[!t]
		\includegraphics[width=1\linewidth]{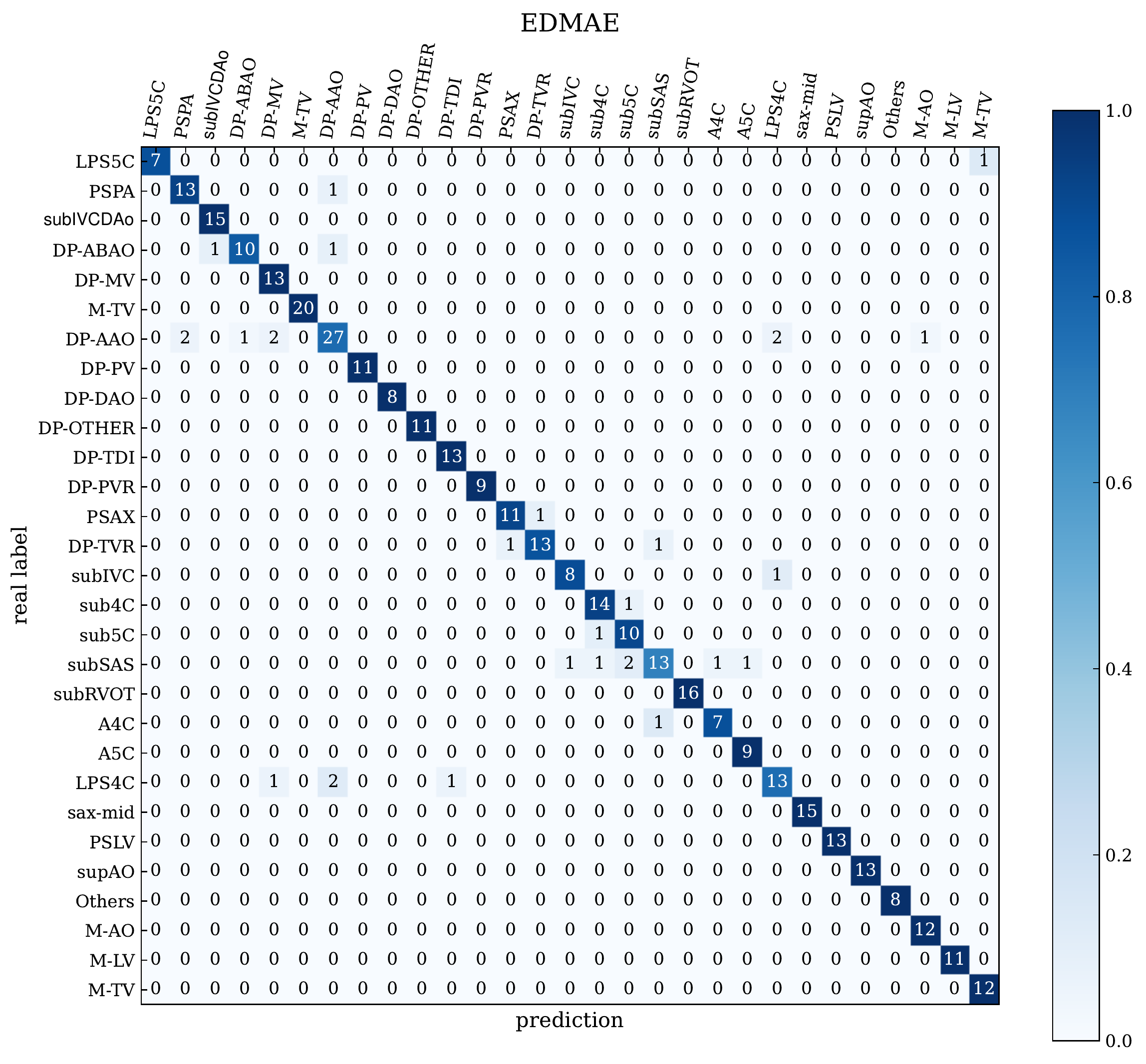}
		\caption{The confusion matrix of the test results of the proposed method on a private dataset.}
		\label{confusion_matrix}
	\end{figure}

	Our dataset consists of 29 categories, which include low parasternal fifive-chamber view (LPS5C), parasternal view of the pulmonary artery (PSPA), parasternal short-axis view (PSAX), parasternal short-axis view at the level of the mitral valve (short axis at mid, sax-mid), parasternal long-axis view of the left ventricle (PSLV), suprasternal long-axis view of the entire aortic arch (supAO), Long axis view of subcostal inferior vena cava (subIVC), subcostal four-chamber view (sub4C), subcostal five-chamber view (sub5C), subcostal sagittal view of the atrium septum (subSAS), subcostal short-axis view through the right ventricular outflflow tract (subRVOT), apical four-chamber view (A4C), apical fifive-chamber view (A5C), low parasternal four-chamber view (LPS4C), transverse section of subxiphoid inferior vena cava and descending aorta (subIVCDAo), other views (others), M-mode echocardiographic recording of the aortic (M-AO), M-mode echocardiography recording of the left ventricle(M-LV), M-mode echocardiography recording of the tricuspid valve (M-TV), Doppler recording from the abdominal aorta (DP-ABAO), Doppler recording from the mitral valve (DP-MV), Doppler recording from the tricuspid valve (DP-MV), Doppler recording from the ascending aorta (DP-AAO), Doppler recording from the pulmonary valve (DP-PV), Doppler recording from the descending aorta (DP-DAO), Doppler recording from the tissue doppler imaging (DP-TDI), other Doppler recordings (DP-OTHER), Doppler recording from the pulmonary valve regurgitation (DP-PVR), Doppler recording from the tricuspid valve regurgitation (DP-TVR) and Doppler recording from the tricuspid valve regurgitation (DP-TVR). As can be seen from Table \ref{results_on_the_private_dataset} and Fig.\ref{confusion_matrix}., the proposed method performs well in classifying most of the views, especially for sub4C, sub5C, and subSAS, which have the best recognition results. However, the recognition performance for other views and DP-OTHER is poor.
	
	\begin{table}[!t]
		\begin{center}
			\caption{Experimental results on the private dataset (the \textcolor{cyan}{\textbf{blue}} value represents the best value in this column, and the \textcolor{red}{red} value represents the worst value in this column).}
			\label{results_on_the_private_dataset}
			\setlength{\tabcolsep}{0.4mm}{
				\begin{tabular}{ l | c | c c c | c }
					\hline
					\makecell[l]{Standard \\ Views} & \makecell[c]{Accuracy \\ (\%)} & \makecell[c]{Precision \\ (\%)} & \makecell[c]{Recall \\ (\%)} & \makecell[c]{Specificity \\ (\%)} & \makecell[c]{F1 \\ (\%)} \\
					\hline
					LPS5C & 99.74 & \textcolor{cyan}{\textbf{99.99}} & 91.66 & \textcolor{cyan}{\textbf{99.99}} & 95.45\\
					PSPA & 99.22 & 85.00 & 95.00 & 99.46 & 89.44\\
					PSAX & 99.74 & 94.44 & \textcolor{cyan}{\textbf{99.99}} & 99.73 & 97.06\\
					subIVCDAo & 99.74 & 90.00 & \textcolor{cyan}{\textbf{99.99}} & 99.73 & 94.44\\
					subIVC & 98.18 & 81.66 & 78.03 & 99.18 & 79.76\\
					sub4C & \textcolor{cyan}{\textbf{99.99}} & \textcolor{cyan}{\textbf{99.99}} & \textcolor{cyan}{\textbf{99.99}} & \textcolor{cyan}{\textbf{99.99}} & \textcolor{cyan}{\textbf{99.99}}\\
					sub5C & \textcolor{cyan}{\textbf{99.99}} & \textcolor{cyan}{\textbf{99.99}} & \textcolor{cyan}{\textbf{99.99}} & \textcolor{cyan}{\textbf{99.99}} & \textcolor{cyan}{\textbf{99.99}}\\
					subSAS & \textcolor{cyan}{\textbf{99.99}} & \textcolor{cyan}{\textbf{99.99}} & \textcolor{cyan}{\textbf{99.99}} & \textcolor{cyan}{\textbf{99.99}} & \textcolor{cyan}{\textbf{99.99}}\\
					subRVOT & \textcolor{cyan}{\textbf{99.99}} & 99.98 & 99.98 & \textcolor{cyan}{\textbf{99.99}} & 99.98\\
					A4C & 99.74 & 93.75 & \textcolor{cyan}{\textbf{99.99}} & 99.73 & 96.66\\
					A5C & \textcolor{cyan}{\textbf{99.99}} & \textcolor{cyan}{\textbf{99.99}} & \textcolor{cyan}{\textbf{99.99}} & \textcolor{cyan}{\textbf{99.99}} & \textcolor{cyan}{\textbf{99.99}}\\
					LPS4C & 99.74 & 92.85 & \textcolor{cyan}{\textbf{99.99}} & 99.73 & 96.15\\
					sax-mid & 99.22 & 93.74 & 88.88 & 99.73 & 91.17\\
					PSLV & 99.22 & 80.91 & \textcolor{cyan}{\textbf{99.99}} & 99.19 & 89.44\\
					supAO & \textcolor{cyan}{\textbf{99.99}} & \textcolor{cyan}{\textbf{99.99}} & \textcolor{cyan}{\textbf{99.99}} & \textcolor{cyan}{\textbf{99.99}} & \textcolor{cyan}{\textbf{99.99}}\\
					Others & \textcolor{red}{\textbf{96.87}} & 87.08 & 77.29 & \textcolor{red}{\textbf{98.85}} & 81.82\\
					M-AO & \textcolor{cyan}{\textbf{99.99}} & \textcolor{cyan}{\textbf{99.99}} & \textcolor{cyan}{\textbf{99.99}} & \textcolor{cyan}{\textbf{99.99}} & \textcolor{cyan}{\textbf{99.99}}\\
					M-LV & \textcolor{cyan}{\textbf{99.99}} & \textcolor{cyan}{\textbf{99.99}} & \textcolor{cyan}{\textbf{99.99}} & \textcolor{cyan}{\textbf{99.99}} & \textcolor{cyan}{\textbf{99.99}}\\
					M-TV & \textcolor{cyan}{\textbf{99.99}} & \textcolor{cyan}{\textbf{99.99}} & \textcolor{cyan}{\textbf{99.99}} & \textcolor{cyan}{\textbf{99.99}} & \textcolor{cyan}{\textbf{99.99}}\\
					DP-ABAO & \textcolor{cyan}{\textbf{99.99}} & \textcolor{cyan}{\textbf{99.99}} & \textcolor{cyan}{\textbf{99.99}} & \textcolor{cyan}{\textbf{99.99}} & \textcolor{cyan}{\textbf{99.99}}\\
					DP-MV & 99.48 & 92.85 & 92.85 & 99.73 & 92.85\\
					DP-TV & 99.22 & 93.75 & 86.60 & 99.73 & 89.90\\
					DP-AAO & 99.48 & 91.66 & 91.66 & 99.73 & 91.66\\
					DP-PV & 99.22 & 87.50 & 92.85 & 99.46 & 90.00\\
					DP-DAO & 98.96 & \textcolor{red}{\textbf{76.19}} & 90.00 & 99.20 & 82.51\\
					DP-TDI & \textcolor{cyan}{\textbf{99.99}} & \textcolor{cyan}{\textbf{99.99}} & \textcolor{cyan}{\textbf{99.99}} & \textcolor{cyan}{\textbf{99.99}} & \textcolor{cyan}{\textbf{99.99}}\\
					DP-PVR & 99.48 & 87.49 & 89.99 & 99.74 & 87.30\\
					DP-TVR & 99.74 & 87.50 & \textcolor{cyan}{\textbf{99.99}} & 99.74 & 92.85\\
					DP-OTHER & 97.92 & 86.60 & \textcolor{red}{\textbf{69.32}} & 99.45 & \textcolor{red}{\textbf{76.84}}\\
					Mean & 99.48 & 93.20 & 94.62 & 99.73 & 93.63\\
					\hline 
			\end{tabular}}
		\end{center}
	\end{table}
	
	\subsection{Experimental results on the public dataset CAMUS}
	To further demonstrate the superiority of the proposed method, a comparison was made with five other methods on the public dataset CAMUS, including MFP-Net~\cite{moradi2019mfp}, Joint-net~\cite{ta2020semi}, TransUNet~\cite{chen2021transunet}, PLANNet~\cite{liu2021deep}, and DenseUNet implemented by ourselves. As shown in Table \ref{results_on_the_public_dataset_CAMUS}, the proposed method outperformed other models in all metrics, with a DC 0.39\% higher than the advanced PLANet and lower HD.
	
\begin{figure}[!t]
		\centering
		\begin{minipage}[b]{1\linewidth}
			\subfloat[Image]{
				\begin{minipage}[b]{0.18\linewidth}
					\centering
					\includegraphics[width=\linewidth]{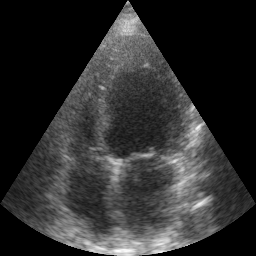}\vspace{4pt}
					\includegraphics[width=\linewidth]{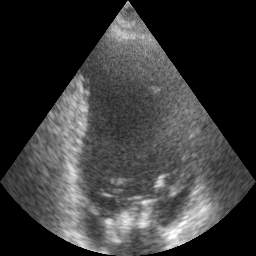}\vspace{4pt}
					\includegraphics[width=\linewidth]{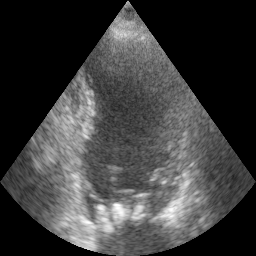}\vspace{4pt}
					\includegraphics[width=\linewidth]{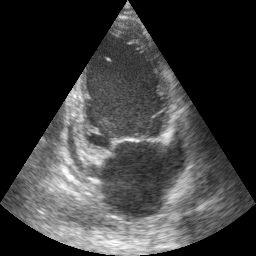}\vspace{4pt}
					\includegraphics[width=\linewidth]{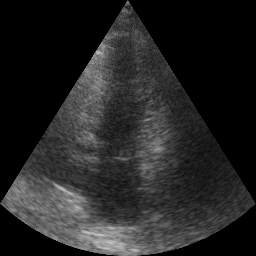}
				\end{minipage}
			}
			\subfloat[ours]{
				\begin{minipage}[b]{0.18\linewidth}
					\centering
					\includegraphics[width=\linewidth]{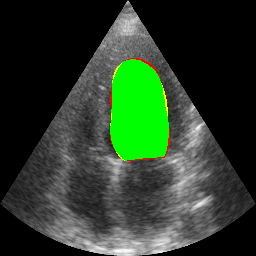}\vspace{4pt}
					\includegraphics[width=\linewidth]{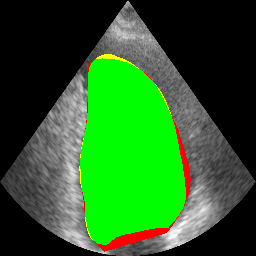}\vspace{4pt}
					\includegraphics[width=\linewidth]{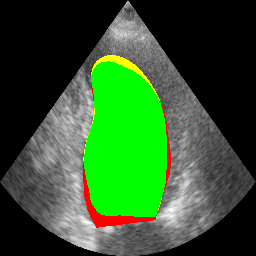}\vspace{4pt}
					\includegraphics[width=\linewidth]{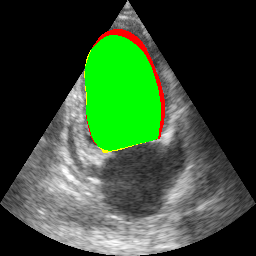}\vspace{4pt}
					\includegraphics[width=\linewidth]{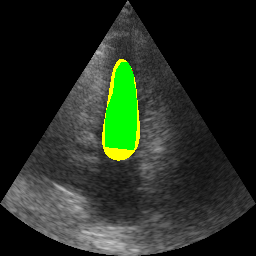}
				\end{minipage}
			}
			\subfloat[PLANet]{
				\begin{minipage}[b]{0.18\linewidth}
					\centering
					\includegraphics[width=\linewidth]{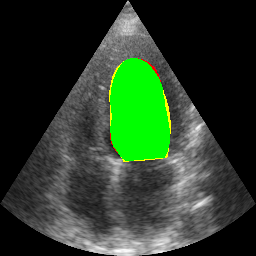}\vspace{4pt}
					\includegraphics[width=\linewidth]{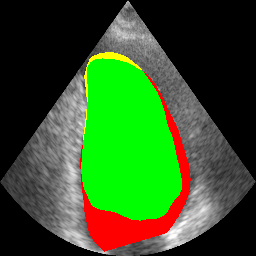}\vspace{4pt}
					\includegraphics[width=\linewidth]{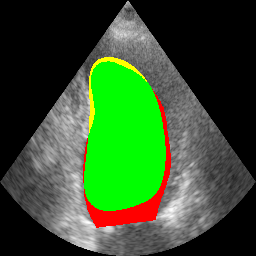}\vspace{4pt}
					\includegraphics[width=\linewidth]{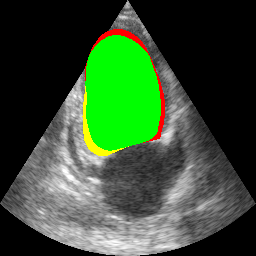}\vspace{4pt}
					\includegraphics[width=\linewidth]{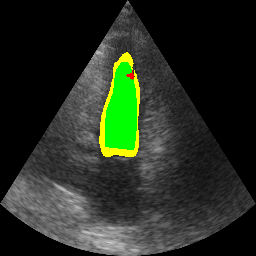}
				\end{minipage}
			}
			\subfloat[TransUNet]{
				\begin{minipage}[b]{0.18\linewidth}
					\centering
					\includegraphics[width=\linewidth]{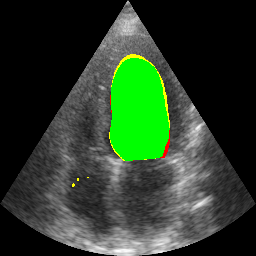}\vspace{4pt}
					\includegraphics[width=\linewidth]{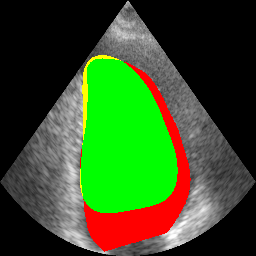}\vspace{4pt}
					\includegraphics[width=\linewidth]{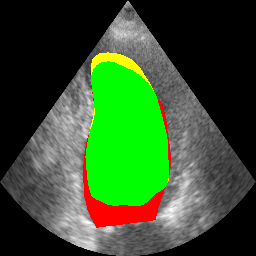}\vspace{4pt}
					\includegraphics[width=\linewidth]{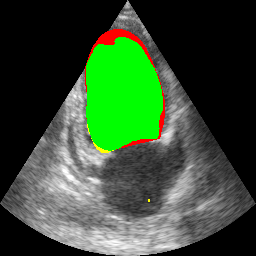}\vspace{4pt}
					\includegraphics[width=\linewidth]{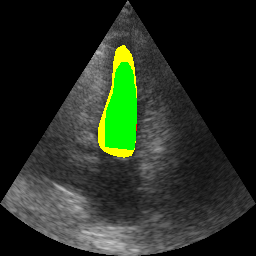}
				\end{minipage}
			}
			\subfloat[DenseUNet]{
				\begin{minipage}[b]{0.18\linewidth}
					\centering
					\includegraphics[width=\linewidth]{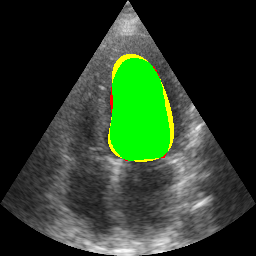}\vspace{4pt}
					\includegraphics[width=\linewidth]{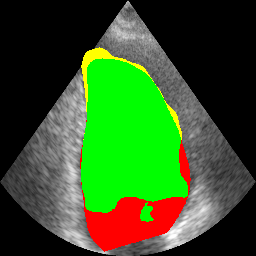}\vspace{4pt}
					\includegraphics[width=\linewidth]{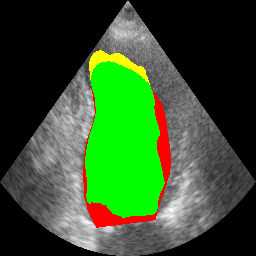}\vspace{4pt}
					\includegraphics[width=\linewidth]{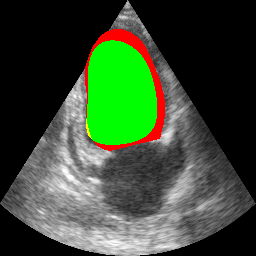}\vspace{4pt}
					\includegraphics[width=\linewidth]{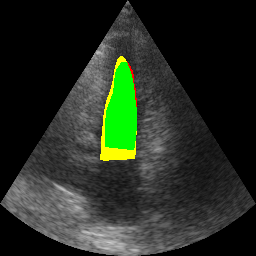}
				\end{minipage}
			}
		\end{minipage}
		\caption{Experimental results on the public dataset CAMUS. The green area represents the overlapping part between the prediction and ground truth, the red area represents the part of ground truth not covered by the prediction, and the yellow area represents the part of the prediction that goes beyond the ground truth.}
		\label{public_dataset_CAMUS_result}
	\end{figure}

	As shown in Fig.\ref{public_dataset_CAMUS_result}., we compared the segmentation results of our proposed method with those of other methods. Our proposed method can achieve good segmentation results on ultrasound images of multiple scales. DenseUNet's segmentation performance is poor, with uneven segmentation edges in large-scale object segmentation and unsatisfactory segmentation results for small-scale objects. However, our DenseUNet model, which underwent self-supervised pretraining, performs much better in segmentation compared to the DenseUNet model without self-supervised pretraining.
	
	\captionsetup[subfloat]{labelsep=none,format=plain,labelformat=empty}
	
	\subsection{Ablation study}
\begin{figure}[htbp]
		\includegraphics[width=1\linewidth]{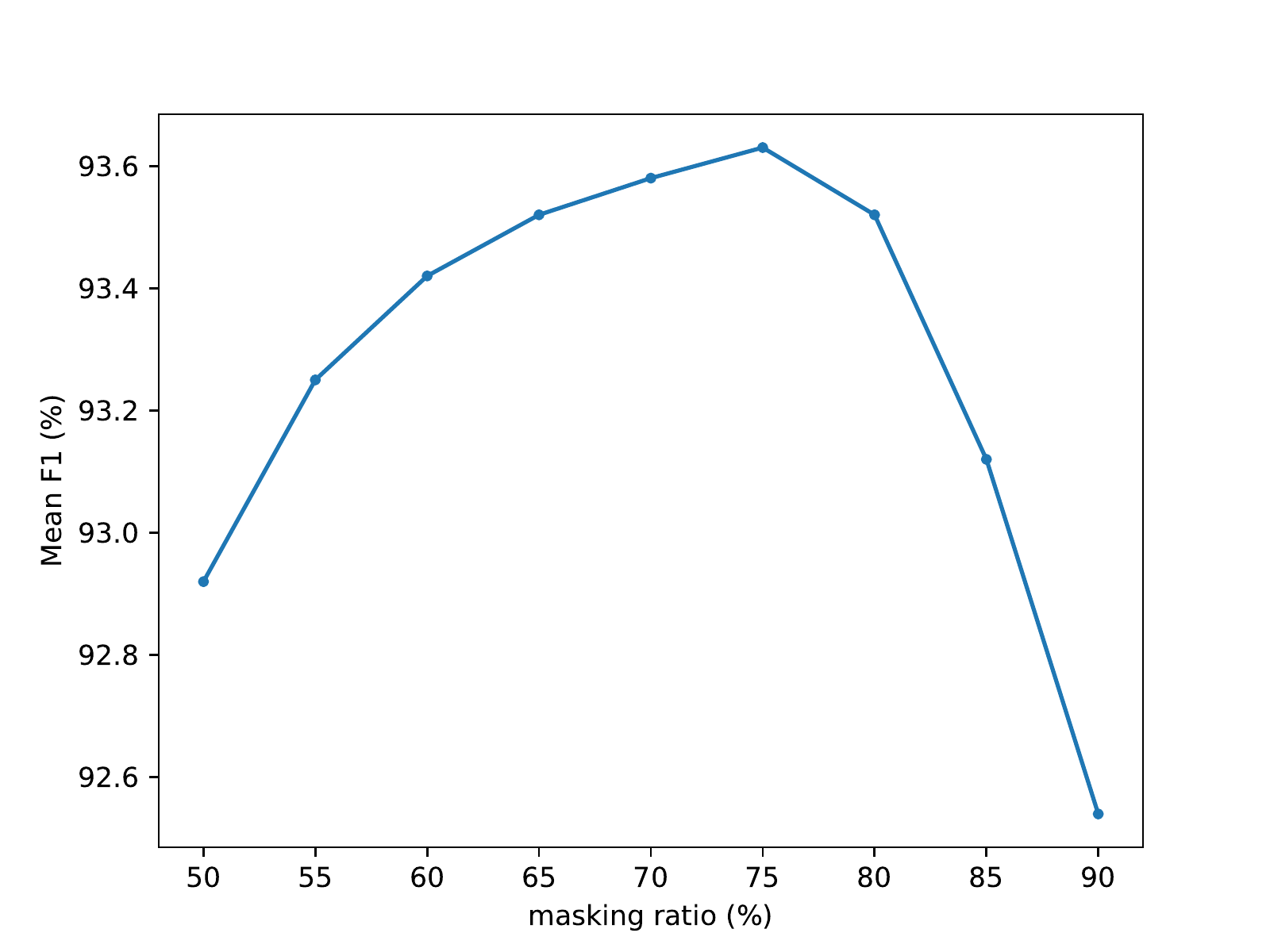}
		\caption{Effect of various masking rates on the performance of standard view recognition in pediatric echocardiography.}
		\label{mask_rate_effect}
	\end{figure}

	The anatomical structures of the heart in the pediatric echocardiograms we collected are relatively fixed, which results in high redundancy in these images. Therefore, employing a higher mask rate for these images can enable the model to learn better potential representations. Although the MAE has demonstrated that a 75\% mask rate is optimal, we have verified this in the task of standard view recognition of pediatric echocardiograms, as shown in Fig.~\ref{mask_rate_effect}.
	
	We compared the convergence time of different self-supervised methods during pre-training on our dataset, as shown in Table~\ref{convergence_time}.
	
	\begin{table}[ht]
		\begin{center}
			\caption{Convergence time of different self-supervised methods.}
			\label{convergence_time}
			\setlength{\tabcolsep}{0.5mm}{
				\begin{tabular}{ l | c | c }
					\hline
					Methods & Mean F1 (\%) & Convergence time (hour)\\
					\hline
					MoCoV2 & 91.85 & 48\\
					MAE & 92.54 & 72\\
					ConvMAE & 92.97 & 48\\
					ours & 93.63 & 24\\
					\hline
			\end{tabular}}
		\end{center}
	\end{table}
	
	Feature alignment is an important step in the proposed method, which aligns the potential representations extracted from the masked image blocks and visible image blocks. This means that the representations obtained by the encoder from any part of the image are consistent, which can force the encoder to learn better representations, while the decoder is only responsible for image reconstruction. Without feature alignment, the encoder may not be fully utilized, which would cause the decoder to learn more representation information, violating the requirements of the decoder's task. In order to verify the impact of feature alignment on the performance of the proposed method, we compared the EDMAE with and without feature alignment, as shown in Table~\ref{feature_alignment_effect}.
	
	\begin{table*}[ht]
		\begin{center}
			\caption{Effect of feature alignment on the performance of Recognition of Standard Views of Pediatric Echocardiography.}
			\label{feature_alignment_effect}
			\setlength{\tabcolsep}{0.5mm}{
				\begin{tabular}{ c | c | c  c  c | c }
					\hline
					Feature Alignment & Overall Accuracy (\%) & Mean Precision (\%) & Mean Recall (\%) & Mean Specificity (\%) & Mean F1 (\%)\\
					\hline
					\XSolidBrush & 98.44 & 92.78 & 93.12 & 99.67 & 92.62\\
					\Checkmark & \textbf{98.48} & \textbf{93.20} & \textbf{94.62} & \textbf{99.73} & \textbf{93.63}\\
					\hline
			\end{tabular}}
		\end{center}
	\end{table*}
	
	The encoder utilized in EDMAE is DenseNet, which possesses a robust fitting ability and suitable parameters and computational complexity. We compared various encoders in EDMAE for verification, which can be seen in Table~\ref{encoder_effect}.
	
	\begin{table*}[ht]
		\begin{center}
			\caption{Effect of various encoders on the performance of Recognition of Standard Views of Pediatric Echocardiography.}
			\label{encoder_effect}
			\setlength{\tabcolsep}{0.5mm}{
				\begin{tabular}{ l | c | c  c  c | c }
					\hline
					Encoder & Overall Accuracy (\%) & Mean Precision (\%) & Mean Recall (\%) & Mean Specificity (\%) & Mean F1 (\%)\\
					\hline
					MobileNetV3 & 98.39 & 91.79 & 93.21 & 99.67 & 92.48\\
					ResNet50 & 98.42 & 91.81 & 93.32 & 99.67 & 92.55\\
					Swin-T-B & 98.44 & 92.79 & 93.14 & 99.68 & 92.53\\
					ours & \textbf{98.48} & \textbf{93.20} & \textbf{94.62} & \textbf{99.73} & \textbf{93.63}\\
					\hline
			\end{tabular}}
		\end{center}
	\end{table*}
	
	As the mainstream classification heads in classification tasks are fully connected layers and rarely use other classification heads, this article will not compare different classification heads. However, there are many types of segmentation heads, and this paper uses the segmentation head of the most classic and simple UNet. We selected some mainstream segmentation heads for comparison, including the segmentation heads of FCN~\cite{long2015fully}, DeepLabV3+~\cite{chen2018encoder}, PSPNet~\cite{zhao2017pyramid}, and OCRNet~\cite{yuan2020object}, to study their impact on model performance. Their backbone networks are all DenseNet pre-trained through self-supervision. As shown in Table~\ref{segmentation_heads}, different segmentation heads have little impact on model performance. Although some indicators of certain segmentation heads surpass the UNet segmentation head we adopted, their cost-effectiveness is not as good as ours, and it is not the focus of this paper.
	
	\begin{table*}[!t]
		\begin{center}
			\caption{Effect of different segmentation heads on model performance.}
			\label{segmentation_heads}
			\setlength{\tabcolsep}{8mm}{
				\begin{tabular}{ l | c  c  c }
					\hline
					Head & DC (\%) & HD (mm) & AUC (\%)\\
					\hline
					FCN~\cite{long2015fully} & 92.96 $\pm$ 0.31 & 3.24 $\pm$ 0.84 & 97.65 $\pm$ 0.26\\
					DeepLabV3+~\cite{chen2018encoder} & 92.94 $\pm$ 0.34 & 3.31 $\pm$ 0.92 & 97.59 $\pm$ 0.28\\
					PSPNet~\cite{zhao2017pyramid} & \textbf{93.12 $\pm$ 0.38} & 3.02 $\pm$ 0.96 & 97.83 $\pm$ 0.22\\
					OCRNet~\cite{yuan2020object} & 93.06 $\pm$ 0.29 & 3.02 $\pm$ 0.97 & \textbf{97.86 $\pm$ 0.24}\\
					ours & 93.09 $\pm$ 0.22 & \textbf{3.02 $\pm$ 0.81} & 97.84 $\pm$ 0.22\\
					\hline 
			\end{tabular}}
		\end{center}
	\end{table*}
	
	In addition, we compared the impact of loss functions on model performance. As can be seen from Table~\ref{loss_function}, the FocalLoss we selected can balance classes well and is more competitive than the classical cross-entropy loss function.
	
	\begin{table*}[!t]
		\begin{center}
			\caption{Effect of different loss functions on model performance.}
			\label{loss_function}
			\setlength{\tabcolsep}{8mm}{
				\begin{tabular}{ l | c  c  c }
					\hline
					Loss Function & DC (\%) & HD (mm) & AUC (\%)\\
					\hline
					Cross Entropy Loss & 92.92 $\pm$ 0.26 & 3.24 $\pm$ 0.89 & 97.75 $\pm$ 0.24\\
					Focal Loss & \textbf{93.09 $\pm$ 0.22} & \textbf{3.02 $\pm$ 0.81} & \textbf{97.84 $\pm$ 0.22}\\
					\hline 
			\end{tabular}}
		\end{center}
	\end{table*}
	
	\section{Discussion}
	EDMAE achieved excellent classification and segmentation performance through self-supervised pre-training on a large-scale dataset of pediatric cardiac ultrasound. From the experiments described above, it can be seen that the proposed method has significant advantages over other methods in downstream tasks such as pediatric cardiac standard view recognition, with good recognition performance for most views and only poor recognition for views with less distinct features. In addition, the proposed method performs well on the public dataset CAMUS and outperforms many advanced methods, showing good performance for object segmentation at multiple scales.
	
	There are three primary factors contributing to the outstanding performance of EDMAE. Firstly, the self-supervised pre-training data distribution is similar to that of downstream tasks, allowing models trained on large-scale data to effectively learn the data distribution. Secondly, the encoder of EDMAE is decoupled from the decoder, which compels the encoder to completely extract the latent semantic representation. Finally, the encoder of the proposed method is a pure convolution operation, which has faster convergence speed and requires less pre-training data.
	
	Although we strive to decouple the encoder and decoder and make them perform their respective duties, this does not mean that the decoder has not learned potential representations, or that we may not have allowed the decoder to fully focus on reconstructing images. In addition, this model is designed for pediatric echocardiography tasks and has not been validated on other types of ultrasound images or other types of medical images. In the future, we will explore new methods to force the encoder-decoder to decouple and perform their respective tasks. In addition, we will extend our approach to multiple ultrasound or medical images to promote and validate our method.

	\section{Conclusion}
	In this paper, an efficient decoupled masked autoencoder with the strong feature extraction ability is proposed for standard view recognition on pediatric echocardiography. The model pre-trained on a private large-scale children's cardiac ultrasound dataset has shown excellent performance in the downstream task of children's heart standard view recognition, which surpasses some advanced classification methods. The proposed model also can be applied in another downstream task, i.e., cardiac ultrasound segmentation, which achieves good segmentation performance. Since the training images are collected from clinical examination database, the proposed method with the high recognition rate for standard view recognition can provide a good technical basis for intelligent diagnosis of congenital heart disease. It would become a new standardized training method for primary-level cardiac ultrasound physicians to practice cardiac view scanning.
	
	\bibliography{references}

\begin{thebibliography}{10}
\expandafter\ifx\csname url\endcsname\relax
  \def\url#1{\texttt{#1}}\fi
\expandafter\ifx\csname urlprefix\endcsname\relax\def\urlprefix{URL }\fi
\expandafter\ifx\csname href\endcsname\relax
  \def\href#1#2{#2} \def\path#1{#1}\fi

\bibitem{zhao2019prevalence}
Q.-M. Zhao, F.~Liu, L.~Wu, X.-J. Ma, C.~Niu, G.-Y. Huang, Prevalence of
  congenital heart disease at live birth in china, The Journal of pediatrics
  204 (2019) 53--58.

\bibitem{douglas2011accf}
P.~S. Douglas, M.~J. Garcia, D.~E. Haines, W.~W. Lai, W.~J. Manning, A.~R.
  Patel, M.~H. Picard, D.~M. Polk, M.~Ragosta, R.~P. Ward, et~al.,
  Accf/ase/aha/asnc/hfsa/hrs/scai/sccm/scct/scmr 2011 appropriate use criteria
  for echocardiography: a report of the american college of cardiology
  foundation appropriate use criteria task force, american society of
  echocardiography, american heart association, american society of nuclear
  cardiology, heart failure society of america, heart rhythm society, society
  for cardiovascular angiography and interventions, society of critical care
  medicine, society of cardiovascular computed tomography, and society for
  cardiovascular magnetic resonance endorsed by the american college of chest
  physicians, Journal of the American College of Cardiology 57~(9) (2011)
  1126--1166.

\bibitem{lopez2010recommendations}
L.~Lopez, S.~D. Colan, P.~C. Frommelt, G.~J. Ensing, K.~Kendall, A.~K.
  Younoszai, W.~W. Lai, T.~Geva, Recommendations for quantification methods
  during the performance of a pediatric echocardiogram: a report from the
  pediatric measurements writing group of the american society of
  echocardiography pediatric and congenital heart disease council, Journal of
  the American Society of Echocardiography 23~(5) (2010) 465--495.

\bibitem{burgos2020evaluation}
X.~P. Burgos-Artizzu, D.~Coronado-Guti{\'e}rrez, B.~Valenzuela-Alcaraz,
  E.~Bonet-Carne, E.~Eixarch, F.~Crispi, E.~Gratac{\'o}s, Evaluation of deep
  convolutional neural networks for automatic classification of common maternal
  fetal ultrasound planes, Scientific Reports 10~(1) (2020) 1--12.

\bibitem{ronneberger2015u}
O.~Ronneberger, P.~Fischer, T.~Brox, U-net: Convolutional networks for
  biomedical image segmentation, in: Medical Image Computing and
  Computer-Assisted Intervention--MICCAI 2015: 18th International Conference,
  Munich, Germany, October 5-9, 2015, Proceedings, Part III 18, Springer, 2015,
  pp. 234--241.

\bibitem{jaiswal2020survey}
A.~Jaiswal, A.~R. Babu, M.~Z. Zadeh, D.~Banerjee, F.~Makedon, A survey on
  contrastive self-supervised learning, Technologies 9~(1) (2020) 2.

\bibitem{zhou2022self}
L.~Zhou, H.~Liu, J.~Bae, J.~He, D.~Samaras, P.~Prasanna, Self pre-training with
  masked autoencoders for medical image analysis, arXiv preprint
  arXiv:2203.05573 (2022).

\bibitem{tian2022unsupervised}
Y.~Tian, G.~Pang, Y.~Liu, C.~Wang, Y.~Chen, F.~Liu, R.~Singh, J.~W. Verjans,
  G.~Carneiro, Unsupervised anomaly detection in medical images with a
  memory-augmented multi-level cross-attentional masked autoencoder, arXiv
  preprint arXiv:2203.11725 (2022).

\bibitem{xiao2023delving}
J.~Xiao, Y.~Bai, A.~Yuille, Z.~Zhou, Delving into masked autoencoders for
  multi-label thorax disease classification, in: Proceedings of the IEEE/CVF
  Winter Conference on Applications of Computer Vision, 2023, pp. 3588--3600.

\bibitem{xu2022swin}
Z.~Xu, Y.~Dai, F.~Liu, W.~Chen, Y.~Liu, L.~Shi, S.~Liu, Y.~Zhou, Swin mae:
  Masked autoencoders for small datasets, arXiv preprint arXiv:2212.13805
  (2022).

\bibitem{chen2022multi}
Z.~Chen, Y.~Du, J.~Hu, Y.~Liu, G.~Li, X.~Wan, T.-H. Chang, Multi-modal masked
  autoencoders for medical vision-and-language pre-training, in: Medical Image
  Computing and Computer Assisted Intervention--MICCAI 2022: 25th International
  Conference, Singapore, September 18--22, 2022, Proceedings, Part V, Springer,
  2022, pp. 679--689.

\bibitem{dosovitskiy2020image}
A.~Dosovitskiy, L.~Beyer, A.~Kolesnikov, D.~Weissenborn, X.~Zhai,
  T.~Unterthiner, M.~Dehghani, M.~Minderer, G.~Heigold, S.~Gelly, et~al., An
  image is worth 16x16 words: Transformers for image recognition at scale,
  arXiv preprint arXiv:2010.11929 (2020).

\bibitem{he2022masked}
K.~He, X.~Chen, S.~Xie, Y.~Li, P.~Doll{\'a}r, R.~Girshick, Masked autoencoders
  are scalable vision learners, in: Proceedings of the IEEE/CVF Conference on
  Computer Vision and Pattern Recognition, 2022, pp. 16000--16009.

\bibitem{liu2021swin}
Z.~Liu, Y.~Lin, Y.~Cao, H.~Hu, Y.~Wei, Z.~Zhang, S.~Lin, B.~Guo, Swin
  transformer: Hierarchical vision transformer using shifted windows, in:
  Proceedings of the IEEE/CVF international conference on computer vision,
  2021, pp. 10012--10022.

\bibitem{huang2017densely}
G.~Huang, Z.~Liu, L.~Van Der~Maaten, K.~Q. Weinberger, Densely connected
  convolutional networks, in: Proceedings of the IEEE conference on computer
  vision and pattern recognition, 2017, pp. 4700--4708.

\bibitem{bao2021beit}
H.~Bao, L.~Dong, S.~Piao, F.~Wei, Beit: Bert pre-training of image
  transformers, arXiv preprint arXiv:2106.08254 (2021).

\bibitem{he2020momentum}
K.~He, H.~Fan, Y.~Wu, S.~Xie, R.~Girshick, Momentum contrast for unsupervised
  visual representation learning, in: Proceedings of the IEEE/CVF conference on
  computer vision and pattern recognition, 2020, pp. 9729--9738.

\bibitem{chen2020simple}
T.~Chen, S.~Kornblith, M.~Norouzi, G.~Hinton, A simple framework for
  contrastive learning of visual representations, in: International conference
  on machine learning, PMLR, 2020, pp. 1597--1607.

\bibitem{goodfellow2020generative}
I.~Goodfellow, J.~Pouget-Abadie, M.~Mirza, B.~Xu, D.~Warde-Farley, S.~Ozair,
  A.~Courville, Y.~Bengio, Generative adversarial networks, Communications of
  the ACM 63~(11) (2020) 139--144.

\bibitem{xie2111simple}
Z.~Xie, Z.~Zhang, Y.~Cao, Y.~Lin, J.~Bao, Z.~Yao, Q.~Dai, H.~S. Hu, A simple
  framework for masked image modeling. arxiv 2021, arXiv preprint
  arXiv:2111.09886.

\bibitem{chen2022context}
X.~Chen, M.~Ding, X.~Wang, Y.~Xin, S.~Mo, Y.~Wang, S.~Han, P.~Luo, G.~Zeng,
  J.~Wang, Context autoencoder for self-supervised representation learning,
  arXiv preprint arXiv:2202.03026 (2022).

\bibitem{fu2022contextual}
Z.~Fu, W.~Zhou, J.~Xu, H.~Zhou, L.~Li, Contextual representation learning
  beyond masked language modeling, arXiv preprint arXiv:2204.04163 (2022).

\bibitem{sowrirajan2021moco}
H.~Sowrirajan, J.~Yang, A.~Y. Ng, P.~Rajpurkar, Moco pretraining improves
  representation and transferability of chest x-ray models, in: Medical Imaging
  with Deep Learning, PMLR, 2021, pp. 728--744.

\bibitem{navarro2022self}
F.~Navarro, C.~Watanabe, S.~Shit, A.~Sekuboyina, J.~C. Peeken, S.~E. Combs,
  B.~H. Menze, Self-supervised pretext tasks in model robustness \&
  generalizability: A revisit from medical imaging perspective, in: 2022 44th
  Annual International Conference of the IEEE Engineering in Medicine \&
  Biology Society (EMBC), IEEE, 2022, pp. 5074--5079.

\bibitem{ly2022student}
S.~T. Ly, B.~Lin, H.~Q. Vo, D.~Maric, B.~Roysam, H.~V. Nguyen, Student
  collaboration improves self-supervised learning: Dual-loss adaptive masked
  autoencoder for brain cell image analysis, arXiv preprint arXiv:2205.05194
  (2022).

\bibitem{quan2022global}
H.~Quan, X.~Li, W.~Chen, M.~Zou, R.~Yang, T.~Zheng, R.~Qi, X.~Gao, X.~Cui,
  Global contrast masked autoencoders are powerful pathological representation
  learners, arXiv preprint arXiv:2205.09048 (2022).

\bibitem{bengio2013representation}
Y.~Bengio, A.~Courville, P.~Vincent, Representation learning: A review and new
  perspectives, IEEE transactions on pattern analysis and machine intelligence
  35~(8) (2013) 1798--1828.

\bibitem{hinton1993autoencoders}
G.~E. Hinton, R.~Zemel, Autoencoders, minimum description length and helmholtz
  free energy, Advances in neural information processing systems 6 (1993).

\bibitem{jiang2022self}
J.~Jiang, N.~Tyagi, K.~Tringale, C.~Crane, H.~Veeraraghavan, Self-supervised 3d
  anatomy segmentation using self-distilled masked image transformer (smit),
  in: Medical Image Computing and Computer Assisted Intervention--MICCAI 2022:
  25th International Conference, Singapore, September 18--22, 2022,
  Proceedings, Part IV, Springer, 2022, pp. 556--566.

\bibitem{howard2019searching}
A.~Howard, M.~Sandler, G.~Chu, L.-C. Chen, B.~Chen, M.~Tan, W.~Wang, Y.~Zhu,
  R.~Pang, V.~Vasudevan, et~al., Searching for mobilenetv3, in: Proceedings of
  the IEEE/CVF international conference on computer vision, 2019, pp.
  1314--1324.

\bibitem{he2016deep}
K.~He, X.~Zhang, S.~Ren, J.~Sun, Deep residual learning for image recognition,
  in: Proceedings of the IEEE conference on computer vision and pattern
  recognition, 2016, pp. 770--778.

\bibitem{chen2020improved}
X.~Chen, H.~Fan, R.~Girshick, K.~He, Improved baselines with momentum
  contrastive learning, arXiv preprint arXiv:2003.04297 (2020).

\bibitem{gao2022convmae}
P.~Gao, T.~Ma, H.~Li, J.~Dai, Y.~Qiao, Convmae: Masked convolution meets masked
  autoencoders, arXiv preprint arXiv:2205.03892 (2022).

\bibitem{leclerc2019deep}
S.~Leclerc, E.~Smistad, J.~Pedrosa, A.~{\O}stvik, F.~Cervenansky, F.~Espinosa,
  T.~Espeland, E.~A.~R. Berg, P.-M. Jodoin, T.~Grenier, et~al., Deep learning
  for segmentation using an open large-scale dataset in 2d echocardiography,
  IEEE transactions on medical imaging 38~(9) (2019) 2198--2210.

\bibitem{loshchilov2017fixing}
I.~Loshchilov, F.~Hutter, Fixing weight decay regularization in adam (2017).

\bibitem{lin2017focal}
T.-Y. Lin, P.~Goyal, R.~Girshick, K.~He, P.~Doll{\'a}r, Focal loss for dense
  object detection, in: Proceedings of the IEEE international conference on
  computer vision, 2017, pp. 2980--2988.

\bibitem{ta2020semi}
K.~Ta, S.~S. Ahn, J.~C. Stendahl, A.~J. Sinusas, J.~S. Duncan, A
  semi-supervised joint network for simultaneous left ventricular motion
  tracking and segmentation in 4d echocardiography, in: Medical Image Computing
  and Computer Assisted Intervention--MICCAI 2020: 23rd International
  Conference, Lima, Peru, October 4--8, 2020, Proceedings, Part VI 23,
  Springer, 2020, pp. 468--477.

\bibitem{chen2021transunet}
J.~Chen, Y.~Lu, Q.~Yu, X.~Luo, E.~Adeli, Y.~Wang, L.~Lu, A.~L. Yuille, Y.~Zhou,
  Transunet: Transformers make strong encoders for medical image segmentation,
  arXiv preprint arXiv:2102.04306 (2021).

\bibitem{moradi2019mfp}
S.~Moradi, M.~G. Oghli, A.~Alizadehasl, I.~Shiri, N.~Oveisi, M.~Oveisi,
  M.~Maleki, J.~Dhooge, Mfp-unet: A novel deep learning based approach for left
  ventricle segmentation in echocardiography, Physica Medica 67 (2019) 58--69.

\bibitem{liu2021deep}
F.~Liu, K.~Wang, D.~Liu, X.~Yang, J.~Tian, Deep pyramid local attention neural
  network for cardiac structure segmentation in two-dimensional
  echocardiography, Medical Image Analysis 67 (2021) 101873.

\bibitem{long2015fully}
J.~Long, E.~Shelhamer, T.~Darrell, Fully convolutional networks for semantic
  segmentation, in: Proceedings of the IEEE conference on computer vision and
  pattern recognition, 2015, pp. 3431--3440.

\bibitem{chen2018encoder}
L.-C. Chen, Y.~Zhu, G.~Papandreou, F.~Schroff, H.~Adam, Encoder-decoder with
  atrous separable convolution for semantic image segmentation, in: Proceedings
  of the European conference on computer vision (ECCV), 2018, pp. 801--818.

\bibitem{zhao2017pyramid}
H.~Zhao, J.~Shi, X.~Qi, X.~Wang, J.~Jia, Pyramid scene parsing network, in:
  Proceedings of the IEEE conference on computer vision and pattern
  recognition, 2017, pp. 2881--2890.

\bibitem{yuan2020object}
Y.~Yuan, X.~Chen, J.~Wang, Object-contextual representations for semantic
  segmentation, in: Computer Vision--ECCV 2020: 16th European Conference,
  Glasgow, UK, August 23--28, 2020, Proceedings, Part VI 16, Springer, 2020,
  pp. 173--190.

\end{thebibliography}
\end{document}